\documentclass[twocolumn,showpacs,aps,prl,preprintnumbers,amsmath,amssymb]{revtex4}

\usepackage{graphicx}
\usepackage{dcolumn}
\usepackage{bm}
\usepackage{epsfig}

\begin{document}
\date{\today}

\title{Ballistic quantum spin Hall state and enhanced edge backscattering \\ in strong magnetic fields}
\author{G. Tkachov $^{1,2}$ and E. M. Hankiewicz $^{2}$}
\affiliation{$^{1}$ Max Planck Institute for the Physics of Complex Systems, Dresden, Germany \\ 
$^{2}$ Institut f\"ur Theoretische Physik und Astrophysik,
Universit\"at W\"urzburg, Germany}


\begin{abstract}
The quantum spin Hall (QSH) state,  observed in a zero magnetic field in HgTe quantum wells,
respects the time-reversal symmetry and is distinct from quantum Hall (QH) states.
We show that the QSH state persists in strong quantizing fields and is identified by  
counter-propagating (helical) edge channels with nonlinear dispersion inside the band gap. 
If the Fermi level is shifted into the Landau-quantized conduction or valence band,
we find a transition between the QSH and QH regimes.
Near the transition the longitudinal conductance of the helical channels is strongly suppressed 
due to the combined effect of the spectrum nonlinearity and enhanced backscattering. 
It shows a power-law decay $B^{-2N}$ with magnetic field $B$, 
determined by the number of backscatterers on the edge, $N$. 
This suggests a rather simple and practical way to probe the quality of recently realized 
quasiballistic QSH devices using magnetoresistance measurements.
\end{abstract}
\maketitle

{\em Introduction}.-
Recently novel two-dimensional (2D) electronic state - quantum spin Hall (QSH) state -
has been theoretically proposed~\cite{Kane05,Bernevig06,Murakami06}
and experimentally realized in HgTe quantum wells (QWs)~\cite{Koenig07,Koenig08,Roth09}.
It originates from spin-orbit band splitting and is characterized by
time-reversal invariant gapless states on sample edges, where electrons with opposite spins
counter-propagate, while the bulk states are fully gapped.
Such (helical) edge channels make the QSH insulators topologically distinct from 
ordinary band insulators, and hold promise
for reversible manipulation of spin-dependent quantum transport.
Similar edge states and transport have been discussed 
in the quantum Hall regime in graphene \cite{Abanin06,Abanin07,McDonald06,Haldane05,Fertig09}.

In experiments on HgTe QWs~\cite{Koenig07,Koenig08}, the QSH regime 
was detected by measuring the longitudinal conductance of two spin channels 
propagating in the same direction on opposite edges of the sample.
This finding was further substantiated by the observed suppression of 
the edge transport in a magnetic field~\cite{Koenig07,Koenig08},
which breaks the time-reversal symmetry of the QSH state, 
thus revealing the helical edge channels.
The magnetoresistance measurements~\cite{Koenig07,Koenig08} and theory~\cite{Maciejko09} 
have so far been done for large disordered samples. 
In view of the progress in the miniaturization of the QSH devices~\cite{Roth09}  
there is an apparent need to investigate the magnetotransport in the {\em ballistic} 
QSH regime, which is the goal of our work.

The interest in the ballistic QSH transport originates from 
the profound difference between the helical edge states, 
characterized by the dissipative longitudinal conductance $\sim 2e^2/h$~\cite{Koenig07,Koenig08,Roth09},  
and dissipationless chiral quantum Hall (QH) channels~\cite{Halperin82,MacDonald84}. 
As a new test to demonstrate this distinction, we propose to measure the longitudinal magnetoconductance 
of a ballistic HgTe QW when its edge spectrum changes from helical to chiral.    
Such a transition is expected when the Fermi level is driven   
from the band gap into the Landau-quantized conduction or valence band 
where a dissipationless QH state sets in. 
The latter is insensitive to the edge backscattering~\cite{Halperin82,MacDonald84}, 
while on the QSH side of the transition we find strong suppression of 
the two-terminal conductance $g$ due to the backscattering in a magnetic field $B$: 
\begin{equation}
g(\epsilon,B)\propto (E_g - |\epsilon|)^{2N}/B^{2N}, \quad |\epsilon| \to E_g.
\label{g_as}
\end{equation}
Here $\epsilon$ indicates the position of the Fermi level 
with respect to the middle of the band gap (equal to $2E_g$). 
This result contrasts the zero-field conductance 
which increases as the Fermi energy is pushed 
into the metallic-type conduction or valence band~\cite{Koenig07,Koenig08,Roth09}. 
Also, unlike the exponential $B$ decay in strongly disordered systems~\cite{Maciejko09}, 
Eq. (\ref{g_as}) describes a power-law magnetoconductance.   

\begin{figure}[b!]
\includegraphics[width=60mm]{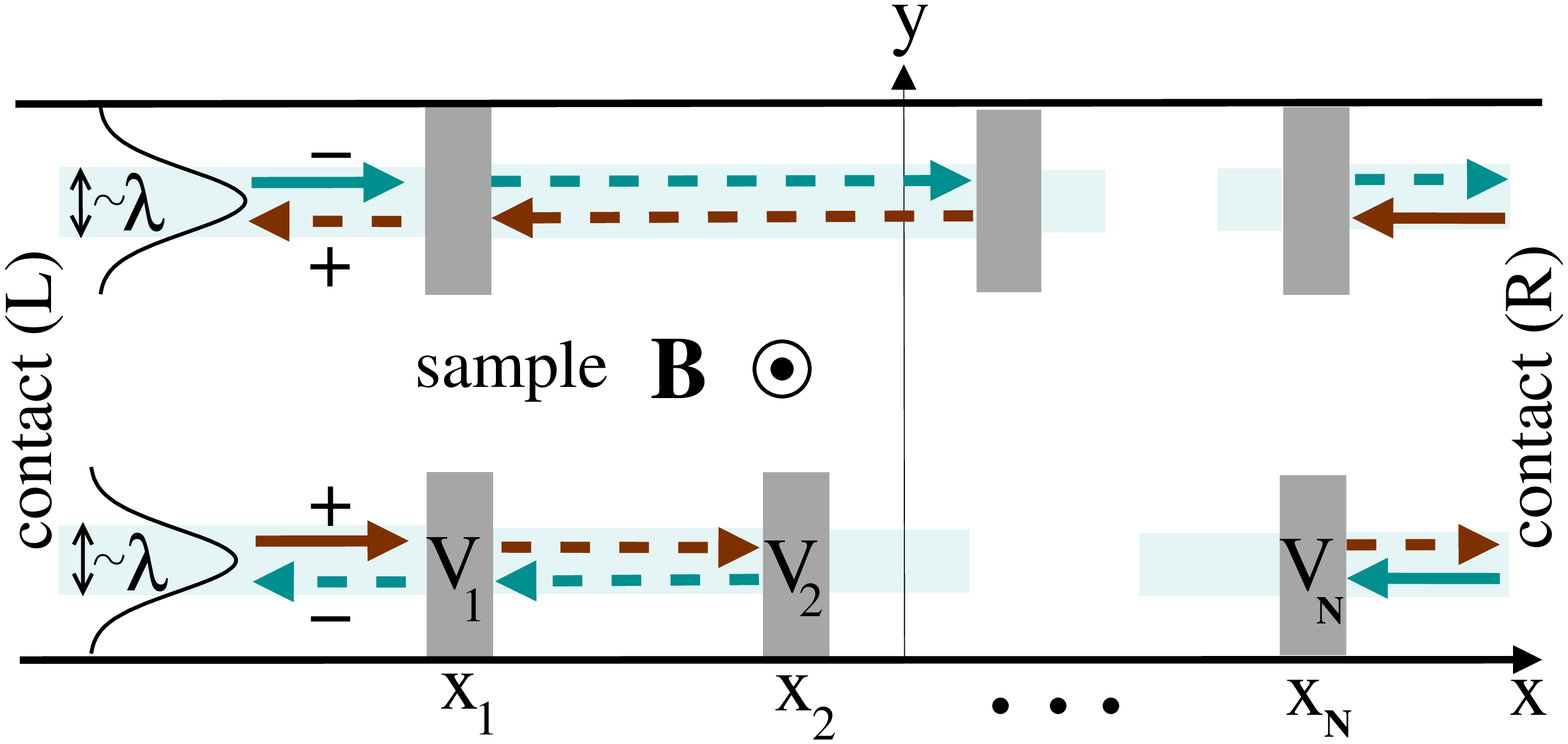}
\caption{(Color online)
Two-terminal QSH device in a strong magnetic field. 
Edge channels are localized within the magnetic length $\lambda$. 
Gray regions schematically indicate backscattering centers (see, also, Eq.~(\ref{V})). 
We assume that the current is carried by the right moving states (solid arrows $\pm$),
populated in contact L and equilibrating in contact R.}
\label{Geo}
\end{figure}

Equation (\ref{g_as}) assumes the presence of a few ($N$) backscattering centers (see, also, Fig.~\ref{Geo}), 
such as sample inhomogeneities where electronic trap states can interact with the edge channels 
randomizing their propagation directions~\cite{Roth09}.     
Although in a zero field this effect is believed to be weak, 
we show that near the QSH-QH transition the backscattering is dramatically enhanced 
due to the reduction of the group velocities of the coupled QSH modes.
According to Eq.~(\ref{g_as}), the analysis of the power of the magnetoconductance decay   
can be a simple tool to determine the quality of the QSH devices, 
which is an important practical task.

{\em Model}.-
We will first analyze the edge states in scattering-free HgTe QWs using 
the effective 4-band model derived in Refs.~\cite{Bernevig06,Koenig08}.
In this approach one works in the basis of the four states near the $\Gamma$  (${\bf k}=0$) 
point of the Brillouin zone:
$|e_1 +\rangle$, $|h_1 +\rangle$, $|e_1 -\rangle$, and $|h_1 -\rangle$,
where $e_1$ and $h_1$ are the s-like electron and p-like hole QW subbands, respectively.  
The index $\tau=\pm$ accounts for the spin degree of freedom.
The effective {\em two-dimensional} Hamiltonian can be approximated by a diagonal matrix in ${\tau}$ space~\cite{Bernevig06,Koenig08}:
\begin{eqnarray}
H=
\left(
  \begin{matrix}
    h_{\bf k} & 0 \\
    0 & h^{\ast}_{\bf -k} \\
  \end{matrix}
\right),
h_{\bf k} = {\bf d}_{\bf k}\mbox{\boldmath$\sigma$},
\,
{\bf d}_{\bf k} =(\hbar\upsilon k_x, -\hbar\upsilon k_y, M).
\label{Heff}
\end{eqnarray}
where Pauli matrices $\sigma_{x,y,z}$ act in subband space,  
$\upsilon\approx 5.5 \times 10^{5} $ms$^{-1}$ is the effective velocity~\cite{Koenig08},
and $M$ determines the band gap $E_g=|M|$ at ${\bf k}=0$.
In Eq.~(\ref{Heff}) we omit terms $\propto {\bf k}^2$ which are small near 
the $\Gamma$ point and in the range of fields we consider~\cite{Trauzettel09}.
We also neglect the bulk inversion asymmetry because we will focus on strong magnetic fields 
where the subband mixing is suppressed.  Up to a unitary transformation, 
Eq.~(\ref{Heff}) is equivalent to a massive Dirac Hamiltonian 
$H_D = \hbar\upsilon \tau_z\mbox{\boldmath$\sigma$}{\bf k} + M\tau_z\sigma_z$ 
[$\tau_z$ is the Pauli matrix in spin space].  
We will work with the corresponding retarded Green's function defined by 
$
[\epsilon\, I  - H_D]{\hat G}({\bf r},{\bf r}^\prime)=I \delta( {\bf r}-{\bf r}^\prime ),
$
where
$
{\bf k}=-i{\bf\nabla}  - e{\bf A}({\bf r})/c\hbar,
$
${\bf A}({\bf r})=(-By,0,0)$ is the vector potential of an external magnetic field $B$, 
and $I=\tau_0\sigma_0={\rm diag}(1,1,1,1)$. 
Assuming a sufficiently wide sample, we find ${\hat G}({\bf r},{\bf r}^\prime)$ near one of the edges,
e.g. $y=0$, using the boundary condition 
${\hat G}({\bf r},{\bf r}^\prime)|_{y=0}=
\tau_0\sigma_x\, {\hat G}({\bf r},{\bf r}^\prime)|_{y=0},
$
equivalent to confinement by infinite "mass" at $y<0$~\cite{Berry87},
\footnote{ This boundary condition can be obtained by introducing a large mass term ($M \to \infty  $) 
outside the physical area of the system~\cite{Berry87}. 
Our results do not strongly depend on the choice of the boundary condition 
since the origin of the QSH edge states is topological: 
a mass domain wall in the inverted regime with $M<0$ in the bulk~\cite{Bernevig06}.
}. 
The matrix ${\hat G}={\rm diag}({\hat G}_+, {\hat G}_-)$ is diagonal in $\tau$ space,
and each ${\hat G}_\tau$ can be diagonalized in e,h space:
\begin{eqnarray}
{\hat G}_\tau=\left(
\begin{array}{cc}
1  & \frac{\upsilon(p_x-ip_y)}{\tau\epsilon - M}\\
\frac{\upsilon(p_x+ip_y)}{\tau\epsilon + M} & 1
\end{array}
\right)
\left(
\begin{array}{cc}
G_{ee|\tau}  & 0\\
0 & G_{hh|\tau}
\end{array}
\right).
\label{G_diag}
\end{eqnarray}
Expanding ${\hat G}$ in plane waves ${\rm e}^{ikx}$ yields 
the boundary problem for the diagonal elements:
\begin{eqnarray}
&
\left[\partial^2_z - \frac{(z - z_k)^2}{4} - a \right]G_{ee|\tau k}=
\frac{\lambda(\epsilon + \tau M)}{\hbar^2\upsilon^2}\delta( z-z^\prime ),
&
\label{Eq1}\\
&
\left.\partial_z G_{ee|\tau k} =
q\, G_{ee|\tau k}\right|_{z=0},
\,
q=\frac{\lambda(\tau \varepsilon + M)}{\hbar \upsilon}  -\lambda k,
&
\label{BC1}
\end{eqnarray}
with $z=y/\lambda$, $z_k=-2\lambda\, k\, {\rm sgn} (eB)$, $\lambda=\sqrt{ c\hbar/2|eB| }$,  and
$a=\lambda^2(M^2 - \epsilon^2)/\hbar^2 \upsilon^2 - \,{\rm sgn} (eB)/2$.
For $G_{hh|\tau k}$ one replaces $\tau, k, B\to  -\tau, -k, -B$.
The solution is obtained in terms of the parabolic cylinder function
$U( a , z)$~\footnote{
$
G_{ee|\tau k}=G^{\infty}_{ee|\tau k}(z,z^\prime) -
C \frac{ \partial_{z_k}U( a , z_k )   + q U( a , z_k )  }
          { \partial_{z_k} U( a ,- z_k ) + q U( a , -z_k ) }\, U( a , z - z_k )U( a , z^\prime - z_k ) .
$
The last term is the edge state, while
$
G^\infty_{ee|\tau k}=C
[
\Theta(z - z^\prime)
U(a,z-z_k)U(a,-z^\prime+z_k)+
\Theta(z^\prime -z)
U(a,z^\prime-z_k)U(a,-z+z_k)
]
$
is the bulk solution with
$
C=-\lambda( \epsilon + \tau M)\Gamma(a +1/2)/\sqrt{2\pi}\hbar^2\upsilon^2$.
For $eB>0$ and using recurrence relations for $U(a,z)$~\cite{AS},
we obtain Eq.~(\ref{G}).
}. It contains the edge contribution of the following form:
\begin{eqnarray}
\hat G_{\tau k}=
\frac{ \alpha(z,z^\prime) \left(
\begin{array}{cc}
1 & \beta(z^\prime)\\
\beta(z) & \beta(z)\beta(z^\prime)
\end{array}
\right) }{\epsilon - \tau M - \tau (\hbar v/\lambda) U(a, -z_k)/U(a+1,-z_k)  },
\label{G}
\end{eqnarray}
\begin{eqnarray}
&
\alpha=\frac{U(a, z-z_k) U(a, z^\prime-z_k) }{\lambda U(a, -z_k)U(a+1,-z_k)},
\beta=\frac{U(a, -z_k) U(a+1, z-z_k) }{U(a+1, -z_k) U(a,z-z_k)}.
&
\label{alpha}
\end{eqnarray}
The new feature of this solution is that it is valid for 
an arbitrary parameter $\hbar\upsilon/\lambda |M|$ 
which measures the magnetic field strength. 
Below we compare {\em weak}- and {\em strong}-field regimes defined by 
$\hbar\upsilon/\lambda |M|\leq 1$ and $\hbar\upsilon/\lambda |M|\gg 1$.

\begin{figure}[t!]
\includegraphics[width=80mm]{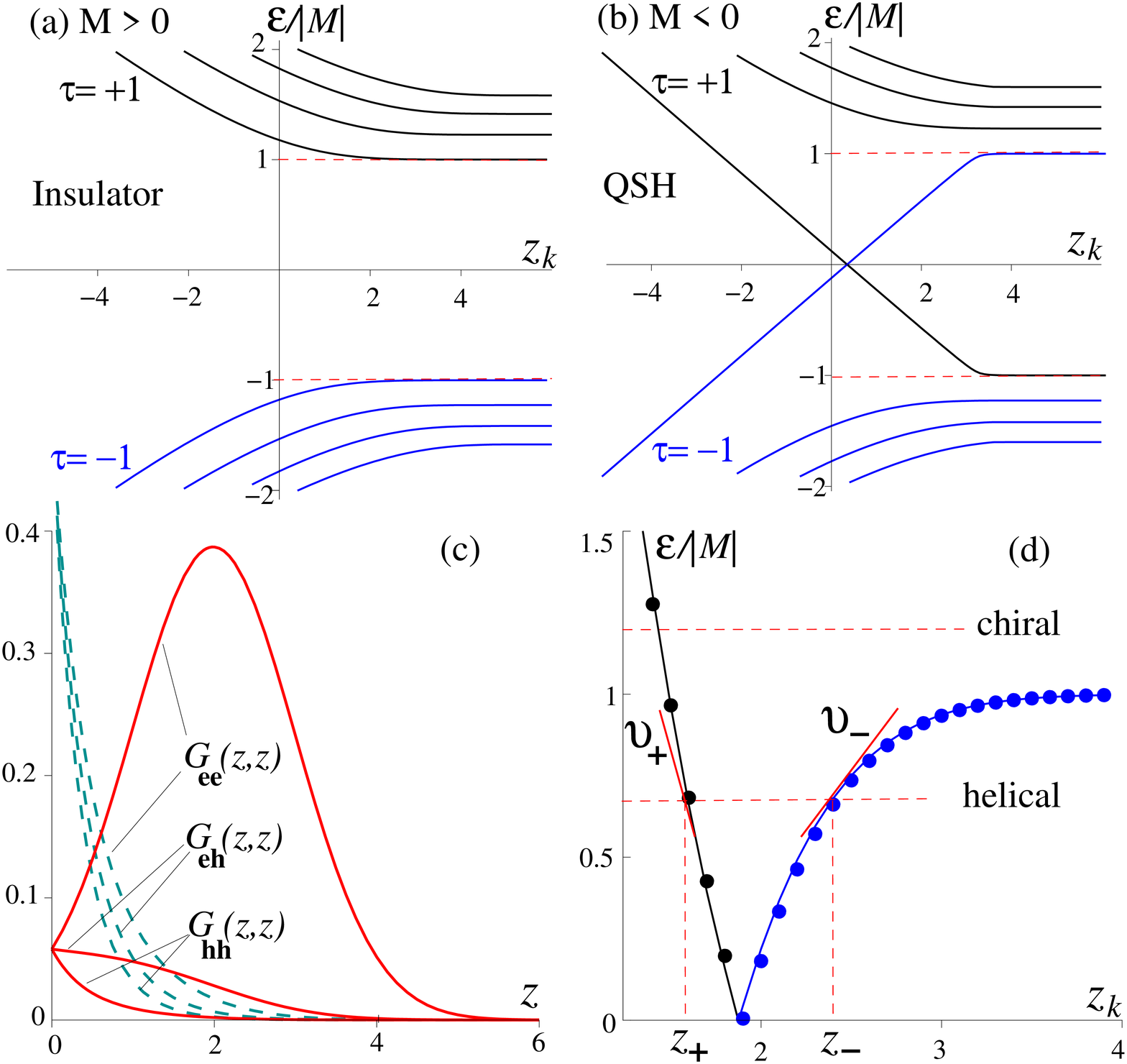}
\caption{(Color online)
Edge-state energy vs. center-of-oscillator coordinate
$z_k=-2\lambda\, k\, {\rm sgn} (eB)$ [$\hbar\upsilon/\lambda |M| =0.5$, $eB>0$]:
(a) insulator and (b) QSH state with two counterpropagating gapless spin channels.
(c) Spatial behavior of zero-energy states in weak (dashed curves)
and strong (solid curves) fields [see, Eq.~(\ref{G}) and text],
(d) Numerical (points) and analytical (Eq.~\ref{E_strong}, solid lines)
results for the spectrum in strong fields, 
$\hbar\upsilon/\lambda |M|\geq 10$. It transforms from helical to chiral at $\epsilon=|M|$. }
\label{Edge}
\end{figure}

{\em Weak- vs. strong-field QSH channels}.-
The edge-state spectrum is given by the pole of Eq.~(\ref{G}).
For weak fields we reproduce the transition from the band insulator with $M>0$ to the QSH state with $M<0$ 
[cf. Figs.~\ref{Edge}(a) and (b)], which is observed at the critical QW thickness $\approx 6.3$ nm~\cite{Bernevig06,Koenig07,Koenig08}.
The QSH state has two gapless counter-propagating spin modes 
exponentially localized at the edge~\cite{Bernevig06,Niu08}, 
as seen from Eq.~(\ref{G}) and Fig.~\ref{Edge}(c) where we use
the asymptotic formula $U(a,z)\approx \sqrt{\pi}/[ 2^{a/2 + 1/4} \Gamma(3/4 + a/2) ]{\rm e}^{-\sqrt{a} \, z}$
with $|a|\gg 1$~\cite{AS}, valid for low fields and energies $|\epsilon| < |M|$:
\begin{eqnarray}
{\hat G}_{\tau k}\approx\frac{(\sigma_0 + \sigma_x)\, \frac{ |M| }{\hbar\upsilon} 
{\rm e}^{ - |M|(y+y^\prime)/\hbar\upsilon }}{ \epsilon  -  \tau M\Theta(M) -  \tau \hbar \upsilon ( k  - k_B)  }, 
\, \frac{\hbar\upsilon}{\lambda  |M| }\ll 1.
\label{G_weak}
\end{eqnarray}
The subgap dispersion is linear: $\epsilon_{\tau k}= \tau \hbar \upsilon ( k  - k_B) $ for $M<0$. 
The magnetic field only shifts the zero-energy point $k_B= -eB \upsilon/(2c|M|)$ with no effect on transport.

As the magnetic field does not open a gap, the QSH state persists in strong fields $\hbar\upsilon/\lambda  |M| \gg 1$,
though the QSH channels are no longer localized at the edge [see, solid curves in Fig.~\ref{Edge}(c)].
The electron function $G_{ee}(z,z)\propto \alpha(z,z)$ for $eB>0$ [or the hole one $G_{hh}(z,z)$ for $eB<0$]
behaves almost like the lowest-Landau-level bulk wave function
peaked at the center of oscillator (COS) $z_k$. The other functions 
are small at $z\sim z_k$. The strong-field asymptotic is obtained for $U(a,z)\approx U(-1/2,z)={\rm e}^{-z^2/4}$,
$U(a+1,z)\approx U(1/2,z)={\rm e}^{z^2/4}\sqrt{\pi/2}\,{\rm erfc}( z/\sqrt{2} )$, and $\beta\ll 1$
in Eqs.~(\ref{G}) and (\ref{alpha}):
\begin{eqnarray}
&&
{\hat G}_{\tau k}\approx  \frac{\sigma_0 + \sigma_z}{2}\,G_{\tau k},\qquad
G_{\tau k}(z,z^\prime)=\frac{\alpha(z,z^\prime)}{ \epsilon - \epsilon_{\tau k} },
\label{G_strong}\\
&&
\alpha(z,z^\prime)\approx \sqrt{ \frac{2}{\pi}  } \frac{ {\rm e}^{ -\frac{(z-z_k)^2}{4} -\frac{(z^\prime-z_k)^2}{4}   } }
{\lambda\, {\rm erfc}( - z_k/\sqrt{2} )},
\label{alpha_strong}\\
&&
\epsilon_{\tau k}= \tau M + \tau \sqrt{ \frac{2}{\pi}  }\, \frac{\hbar\upsilon}{\lambda}\frac{ {\rm e}^{-z_k^2/2 } }
{ {\rm erfc}( - z_k/\sqrt{2} ) },\quad
\frac{\hbar\upsilon}{\lambda  |M| }\gg 1,\quad\quad
\label{E_strong}
\end{eqnarray}
where ${\rm erfc}( z )$ is the complementary error function. 
However, the most essential distinction of this regime is the {\em nonlinear} spectrum (\ref{E_strong}). 
Upon crossing the gap energy $E_g=|M|$ it changes from helical to chiral, 
as illustrated in Fig.~\ref{Edge}(d). Therefore, the QSH state transforms into  
a dissipationless $\nu =1$ QH state~\cite{Halperin82,MacDonald84}.  
Unlike related work on HgTe QWs~\cite{Koenig08,Trauzettel09,Akhmerov09} and graphene~\cite{Abanin06,Fertig09}, 
we intend to study the QSH-QH transition in the energy (e.g. gate voltage) 
dependence of the longitudinal conductance. 
For that purpose, we need the group velocities, $\upsilon_\pm(\epsilon,B)$ and COS coordinates, 
$z_\pm(\epsilon,B)$, which are obtained from Eq.~(\ref{E_strong}) linearized near given energy,
$\epsilon_{\tau k}\approx \epsilon - (\hbar\upsilon_\tau/2\lambda) (z_k - z_\tau)$.
Here $z_{\tau}(\epsilon,B)$ is the solution of equation $\epsilon_{\tau k} =\epsilon$, 
which is related to the velocity by
\begin{eqnarray}
\upsilon_\tau(\epsilon,B)=2\lambda (\tau |M| +\epsilon)\, z_{\tau}(\epsilon,B)/\hbar.
\label{zv}
\end{eqnarray}
The edge state can be described by the one-dimensional Green's function,
$G_\tau (x,x^\prime)=\int \frac{dk}{2\pi}\,e^{i k (x-x^\prime)}$ $\times\int_0^{\infty}dy\, G_{\tau k}(y,y)$,
where $G_{\tau k}(y,y)$ is localized within $\lambda$ 
[see, Eq.~(\ref{alpha_strong})]. Using the linearized dispersion we find
 \begin{eqnarray}
G_{\tau}(x,x^\prime)=
\frac{ {\rm e}^{ ik_\tau (\epsilon,B) (x-x^\prime) }  }
{i\hbar |\upsilon_\tau(\epsilon,B)|}\Theta([x-x^\prime] \tau),
k_\tau=-\frac{z_\tau}{2\lambda}.
\label{G_x}
\end{eqnarray}
The step function $\Theta([x-x^\prime] \tau)$ accounts for the chirality.

{\em Edge backscattering and magnetoconductance}.-
We now calculate the two-terminal conductance of a QSH system [see, Fig.~\ref{Geo}] 
using the scattering matrix formalism.
Since the edges are assumed decoupled, it is sufficient to do the calculations 
for one of them, e.g., for the lower edge in Fig.~\ref{Geo} 
which is described by the following ${\hat S}$ matrix:   
$
{\hat S}=
\bigl(
  \begin{smallmatrix}
   r^{-+}_{_{LL}}  & t^{--}_{_{ LR }} \\
   t^{++}_{_{ RL }}  &  r^{+-}_{_{ RR }} \\
  \end{smallmatrix}
\bigr)\otimes\frac{\sigma_0+\sigma_z}{2}.
$
Here $r$'s and $t$'s are the reflection and transmission amplitudes
for the right ("+")- and left ("-") -moving states;
$(\sigma_0+\sigma_z)/2$ projects the ${\hat S}$ matrix on the electron QW subband
which has the non-vanishing wave function [see, Eq.~(\ref{G_strong})].
The conductance,
$
 g=(e^2/h)\left| t^{++}_{_{RL}}\right|^2,
$
is calculated using Fisher-Lee relation~\cite{FisherLee81}, 
$
t^{++}_{_{RL}}=i\hbar |\upsilon_+|{\cal G}_{++}(x\in R, x^\prime\in L),
$
between $t^{++}_{_{RL}}$ and the diagonal element ${\cal G}_{++}(x,x^\prime)$ of the Green's function 
${\cal \hat G}(x,x^\prime)=
\bigl(
  \begin{smallmatrix}
   {\cal G}_{++}  &  {\cal G}_{+-}\\
   {\cal G}_{-+}  & {\cal G}_{--} 
  \end{smallmatrix}
\bigr).
$
Its {\em off-diagonal} part is due to backscattering.
We model it by the sum of $N$ potentials, localized at positions $x_n$ with  
non-zero matrix elements $V_n$ between the right- and left-moving states:   
\begin{eqnarray}
{\hat V}(x)=\sum\nolimits_{n=1..N} V_n \delta(x-x_n)\, \tau_x.
\label{V}
\end{eqnarray}
Microscopically, the coupling between the counter-propagating channels   
can be mediated by interaction with electronic trap states which are likely to exist 
even in high quality samples~\cite{Roth09}. 
Note that choosing the other off-diagonal matrix, $\tau_y$ does not change the final result. 
Potential~(\ref{V}) results in the Dyson equation 
${\cal\hat G}(x,x^\prime) = {\hat G}(x,x^\prime) +
\sum_{n=1..N}{\hat G}(x,x_n)V_n\,\tau_x\,
{\cal \hat G}(x_n,x^\prime)$.
This allows us to express the off-diagonal part ${\cal G}_{\tau,-\tau}(x,x^\prime)$ through 
${\cal G}_{\tau\tau}(x,x^\prime)$ and obtain a closed equation for the latter: 
$
{\cal G}_{\tau\tau}(x,x^\prime) = G_\tau(x,x^\prime)+
+
\sum_{n,m=1..N}G_\tau(x,x_n)V_nG_{-\tau}(x_n,x_m)V_m  
{\cal G}_{\tau\tau}(x_m,x^\prime).
$
With known unperturbed function $G_\tau$ (\ref{G_x}) and for not large $N$, 
we solve this equation and calculate $g$.  
Let us look first at the particular cases $N=1,2$ and $3$:
\begin{eqnarray}
&
g=\frac{e^2}{h}
\bigl( 1+\frac{V_1^2}{\hbar^2|\upsilon_+\upsilon_-| } \bigr)^{-2},
&
\label{g1}\\
&
g=\frac{e^2}{h}
\bigl| 1 + \frac{V^2_1+V^2_2+V_1V_2{\rm e}^{ iQd_{12} } }{\hbar^2|\upsilon_+\upsilon_-| }
         + \frac{ V^2_1V^2_2 }{\hbar^4\upsilon^2_+\upsilon^2_- }
\bigr|^{-2}
,
&
\label{g2}\\
&
g=\frac{e^2}{h}
\Bigl|1 +\frac{\sum\limits_{n=1}^3V^2_n+V_1V_2{\rm e}^{iQd_{12}}  + V_1V_3{\rm e}^{iQd_{13}} +V_2V_3{\rm e}^{iQd_{23}}   }{\hbar^2|\upsilon_+\upsilon_-| }
&
\nonumber\\
&
         +\frac{ V_1^2V_2^2 + V_1^2V_3^2 + V_2^2V_3^2 + V_1^2V_2V_3{\rm e}^{ iQd_{23} } }{\hbar^4\upsilon^2_+\upsilon^2_- }
         +\frac{ V_1^2V_2^2V_3^2 }{\hbar^6|\upsilon_+\upsilon_-|^3}
\Bigr|^{-2},
&
\label{g3}
\end{eqnarray}
where $Q=k_+ - k_-$ and $d_{nm}=x_m-x_n$. 

It is clear from Eqs.~(\ref{g1}) -- (\ref{g3}) that for an arbitrary $N$ the conductance contains 
the cross product $V_1^2\cdots V_N^2/|\upsilon_+\upsilon_-|^N$ arising from the simultaneous scattering 
from $N$ potentials. This is the most divergent term when one of the velocities $\upsilon_\pm$ 
vanishes near the band gap, $|\epsilon|\to E_g=|M|$  (e.g. $\upsilon_-\to 0$ in Fig.~\ref{Edge}(d)). 
Such strong enhancement of the backscattering leads to the suppressed conductance, 
\begin{eqnarray}
g\approx (e^2/h)\times\hbar^{4N}|\upsilon_+\upsilon_-|^{2N}/(V_1\cdots V_N)^4\ll e^2/h. 
\label{gN}
\end{eqnarray}
Using Eq.~(\ref{zv}) for $\upsilon_\pm$ and ommiting the slower functions $z_\pm(\epsilon,B)$, 
we obtain the qualitative energy and field dependence of 
the conductance near the QSH-QH transition, presented in the introduction [see, Eq.~(\ref{g_as})]. 

In a wider range of energies and fields the typical behavior of the conductance can be understood 
from Eq.~(\ref{g2}) assuming two backscattering centers on the edge.   
First of all, it is easy to verify that Eq.~(\ref{g2}) is valid not only for strong fields,
but also in the weak-field case where the unperturbed Green's function is given by Eq.~(\ref{G_weak}).
Since the weak-field spectrum is linear $\epsilon_{\tau k}= \tau \hbar \upsilon ( k  + k_B) $, 
we have $\upsilon_+=-\upsilon_-=\upsilon$, $k_{\pm }= - k_B \pm \epsilon/\hbar\upsilon $ and $Q=k_+-k_-=2\epsilon/\hbar\upsilon$. Therefore, $g$ is independent of the magnetic field 
and for $V_{1,2}\ll \hbar\upsilon$ is almost independent of energy (see dashed curve in Fig.~3a).
Thus weak channel mixing is hardly detectable for small $B$.
\begin{figure}[t!]
\includegraphics[width=60mm]{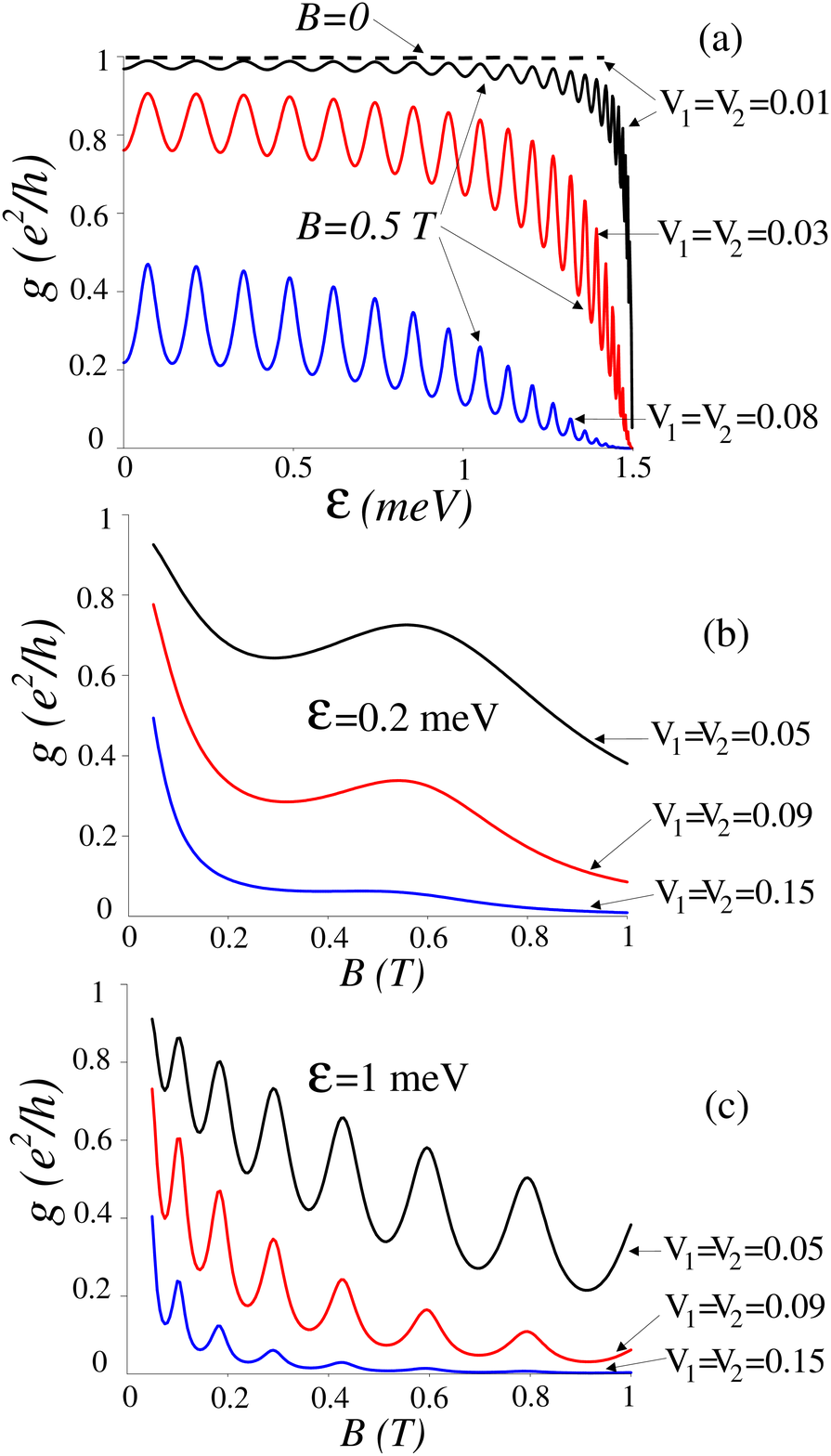}
\caption{(Color online) Conductance [see, Eq.~(\ref{g2})] vs. (a) energy below the band gap $E_g=1.5$ meV and  
(b,c) vs. magnetic field. $V_{1,2}$ are in units of meV$\cdot\mu$m; $d_{12}=3 \mu$m.}
\label{g_fig}
\end{figure}
In contrast, in strong magnetic fields, scattering of the same strength is sufficient to 
suppress the conductance, $g$ near the band gap where the QSH-QH transition occurs  
[cf. dashed and solid curves for $V_{1,2}=0.01$ meV$\cdot\mu$m near $E_g=1.5$ meV in Fig.~\ref{g_fig}(a)].
Figs~\ref{g_fig}(a),(b) and (c) also show that the conductance suppression 
is accompanied by Fabry-Perot-type oscillations due to interference of the counter-propagating channels 
which acquire the energy- and field-dependent 
phase difference $(k_+ - k_-)d_{12}$, in scattering between the defects.

In our model the upper magnetic field limit lies in the range of a few Tesla. 
This estimate is based on Refs.~\cite{Koenig08,Trauzettel09} predicting another ($B$-field induced) 
QSH-QH transition due to a quadratic correction ${\cal B}{\bf k}^2$ to the mass term in effective Hamiltonian~(\ref{Heff}). The smallness of the parameter ${\cal B}|M|/2\hbar^2\upsilon^2\ll 1$~\cite{Koenig08,Trauzettel09} allows us to neglect such ${\bf k}^2$ term and to meet, at the same time, 
the strong field condition $\hbar\upsilon/\lambda |M|> 1$.

{\em Conclusions}.- We have studied the longitudinal conductance of helical spin edge channels 
in HgTe quantum wells in strong magnetic fields. As the Fermi level approaches the band gap, 
the conductance vanishes due to the combined effect of the spectral nonlinearity and channel backscattering. 
Such energy dependence indicates a transition between the quantum spin Hall 
and dissipationless quantum Hall regimes.
The conductance exhibits $B^{-2N}$ magnetic field dependence, determined by the number $N$ of 
backscattering centers on the edge. This suggests a simple way to detect defects in ballistic QSH devices 
using standard magnetoresistance measurements.

{\em Acknowledgements.}
We thank Shou-Cheng Zhang, Qiaoliang Qi, Joseph Maciejko, Alena Novik, Hartmut Buhmann, Laurens Molenkamp 
and Bj\"orn Trauzettel for enlightening discussions.
The work was financially supported by the DFG Emmy-Noether Programme (G.T.) and
by DFG grant HA5893/1-1 (G.T. and E.H.).


\begin{thebibliography}{20}
\expandafter\ifx\csname natexlab\endcsname\relax\def\natexlab#1{#1}\fi
\expandafter\ifx\csname bibnamefont\endcsname\relax
  \def\bibnamefont#1{#1}\fi
\expandafter\ifx\csname bibfnamefont\endcsname\relax
  \def\bibfnamefont#1{#1}\fi
\expandafter\ifx\csname citenamefont\endcsname\relax
  \def\citenamefont#1{#1}\fi
\expandafter\ifx\csname url\endcsname\relax
  \def\url#1{\texttt{#1}}\fi
\expandafter\ifx\csname urlprefix\endcsname\relax\def\urlprefix{URL }\fi
\providecommand{\bibinfo}[2]{#2}
\providecommand{\eprint}[2][]{\url{#2}}

\bibitem[{\citenamefont{Kane and Mele}(2005)}]{Kane05}
\bibinfo{author}{\bibfnamefont{C.~L.} \bibnamefont{Kane}} \bibnamefont{and}
  \bibinfo{author}{\bibfnamefont{E.~J.} \bibnamefont{Mele}},
  \bibinfo{journal}{Phys. Rev. Lett.} \textbf{\bibinfo{volume}{95}},
  \bibinfo{pages}{226801} (\bibinfo{year}{2005}).

\bibitem[{\citenamefont{Bernevig et~al.}(2006)\citenamefont{Bernevig, Hughes,
  and Zhang}}]{Bernevig06}
\bibinfo{author}{\bibfnamefont{B.~A.} \bibnamefont{Bernevig}},
  \bibinfo{author}{\bibfnamefont{T.~L.} \bibnamefont{Hughes}},
  \bibnamefont{and} \bibinfo{author}{\bibfnamefont{S.~C.} \bibnamefont{Zhang}},
  \bibinfo{journal}{Science} \textbf{\bibinfo{volume}{314}},
  \bibinfo{pages}{1757} (\bibinfo{year}{2006}).

\bibitem[{\citenamefont{Murakami}(2006)}]{Murakami06}
\bibinfo{author}{\bibfnamefont{S.}~\bibnamefont{Murakami}},
  \bibinfo{journal}{Phys. Rev. Lett.} \textbf{\bibinfo{volume}{97}},
  \bibinfo{pages}{236805} (\bibinfo{year}{2006}).

\bibitem[{\citenamefont{K{\"o}nig et~al.}(2007)}]{Koenig07}
\bibinfo{author}{\bibfnamefont{M.}~\bibnamefont{K{\"o}nig}}
  \bibnamefont{et~al.}, \bibinfo{journal}{Science}
  \textbf{\bibinfo{volume}{318}}, \bibinfo{pages}{766} (\bibinfo{year}{2007}).

\bibitem[{\citenamefont{K{\"o}nig et~al.}(2008)}]{Koenig08}
\bibinfo{author}{\bibfnamefont{M.}~\bibnamefont{K{\"o}nig}}
  \bibnamefont{et~al.}, \bibinfo{journal}{J. Phys. Soc. Jpn.}
  \textbf{\bibinfo{volume}{77}}, \bibinfo{pages}{031007}
  (\bibinfo{year}{2008}).

\bibitem[{\citenamefont{Roth et~al.}(2009)}]{Roth09}
\bibinfo{author}{\bibfnamefont{A.}~\bibnamefont{Roth}} \bibnamefont{et~al.},
  \bibinfo{journal}{Science} \textbf{\bibinfo{volume}{325}},
  \bibinfo{pages}{294} (\bibinfo{year}{2009}).

\bibitem[{\citenamefont{Abanin et~al.}(2006)\citenamefont{Abanin, Lee, and
  Levitov}}]{Abanin06}
\bibinfo{author}{\bibfnamefont{D.~A.} \bibnamefont{Abanin}},
  \bibinfo{author}{\bibfnamefont{P.~A.} \bibnamefont{Lee}}, \bibnamefont{and}
  \bibinfo{author}{\bibfnamefont{L.~S.} \bibnamefont{Levitov}},
  \bibinfo{journal}{Phys. Rev. Lett.} \textbf{\bibinfo{volume}{96}},
  \bibinfo{pages}{176803} (\bibinfo{year}{2006}).

\bibitem[{\citenamefont{Abanin et~al.}(2007)}]{Abanin07}
\bibinfo{author}{\bibfnamefont{D.~A.} \bibnamefont{Abanin}}
  \bibnamefont{et~al.}, \bibinfo{journal}{Phys. Rev. Lett.}
  \textbf{\bibinfo{volume}{98}}, \bibinfo{pages}{196806}
  (\bibinfo{year}{2007}).

\bibitem[{\citenamefont{Nomura and MacDonald}(2006)}]{McDonald06}
\bibinfo{author}{\bibfnamefont{K.}~\bibnamefont{Nomura}} \bibnamefont{and}
  \bibinfo{author}{\bibfnamefont{A.~H.} \bibnamefont{MacDonald}},
  \bibinfo{journal}{Phys. Rev. Lett.} \textbf{\bibinfo{volume}{96}},
  \bibinfo{pages}{256602} (\bibinfo{year}{2006}).

\bibitem[{\citenamefont{Sheng et~al.}(2005)}]{Haldane05}
\bibinfo{author}{\bibfnamefont{L.}~\bibnamefont{Sheng}} \bibnamefont{et~al.},
  \bibinfo{journal}{Phys. Rev. Lett.} \textbf{\bibinfo{volume}{95}},
  \bibinfo{pages}{136602} (\bibinfo{year}{2005}).

\bibitem[{\citenamefont{Shimshoni et~al.}(2009)\citenamefont{Shimshoni, Fertig,
  and Pai}}]{Fertig09}
\bibinfo{author}{\bibfnamefont{E.}~\bibnamefont{Shimshoni}},
  \bibinfo{author}{\bibfnamefont{H.~A.} \bibnamefont{Fertig}},
  \bibnamefont{and} \bibinfo{author}{\bibfnamefont{G.~V.} \bibnamefont{Pai}},
  \bibinfo{journal}{Phys. Rev. Lett.} \textbf{\bibinfo{volume}{102}},
  \bibinfo{pages}{206408} (\bibinfo{year}{2009}).

\bibitem[{\citenamefont{Maciejko et~al.}()\citenamefont{Maciejko, Qi, and
  Zhang}}]{Maciejko09}
\bibinfo{author}{\bibfnamefont{J.}~\bibnamefont{Maciejko}},
  \bibinfo{author}{\bibfnamefont{X.-L.} \bibnamefont{Qi}}, \bibnamefont{and}
  \bibinfo{author}{\bibfnamefont{S.-C.} \bibnamefont{Zhang}},
  \bibinfo{howpublished}{arXiv:0907.4515}.

\bibitem[{\citenamefont{Halperin}(1982)}]{Halperin82}
\bibinfo{author}{\bibfnamefont{B.~I.} \bibnamefont{Halperin}},
  \bibinfo{journal}{Phys. Rev. B} \textbf{\bibinfo{volume}{25}},
  \bibinfo{pages}{2185} (\bibinfo{year}{1982}).

\bibitem[{\citenamefont{MacDonald and Streda}(1984)}]{MacDonald84}
\bibinfo{author}{\bibfnamefont{A.~H.} \bibnamefont{MacDonald}}
  \bibnamefont{and} \bibinfo{author}{\bibfnamefont{P.}~\bibnamefont{Streda}},
  \bibinfo{journal}{Phys. Rev. B} \textbf{\bibinfo{volume}{29}},
  \bibinfo{pages}{1616} (\bibinfo{year}{1984}).

\bibitem[{\citenamefont{Schmidt et~al.}(2009)}]{Trauzettel09}
\bibinfo{author}{\bibfnamefont{M.~J.} \bibnamefont{Schmidt}}
  \bibnamefont{et~al.}, \bibinfo{journal}{Phys. Rev. B}
  \textbf{\bibinfo{volume}{79}}, \bibinfo{pages}{241306(R)}
  (\bibinfo{year}{2009}).

\bibitem[{\citenamefont{Berry and Mondragon}(1987)}]{Berry87}
\bibinfo{author}{\bibfnamefont{M.~V.} \bibnamefont{Berry}} \bibnamefont{and}
  \bibinfo{author}{\bibfnamefont{R.~J.} \bibnamefont{Mondragon}},
  \bibinfo{journal}{Proc. R. Soc. Lond. A} \textbf{\bibinfo{volume}{412}},
  \bibinfo{pages}{53} (\bibinfo{year}{1987}).

\bibitem[{\citenamefont{Zhou et~al.}(2008)}]{Niu08}
\bibinfo{author}{\bibfnamefont{B.}~\bibnamefont{Zhou}} \bibnamefont{et~al.},
  \bibinfo{journal}{Phys. Rev. Lett.} \textbf{\bibinfo{volume}{101}},
  \bibinfo{pages}{246807} (\bibinfo{year}{2008}).

\bibitem[{\citenamefont{Abramowitz and Stegun}(1964)}]{AS}
\bibinfo{author}{\bibfnamefont{M.}~\bibnamefont{Abramowitz}} \bibnamefont{and}
  \bibinfo{author}{\bibfnamefont{I.}~\bibnamefont{Stegun}},
  \emph{\bibinfo{title}{Handbook of Mathematical Functions with Formulas,
  Graphs, and Mathematical Tables}} (\bibinfo{publisher}{National Bureau of
  Standards}, \bibinfo{year}{1964}).

\bibitem[{\citenamefont{Akhmerov et~al.}(2009)}]{Akhmerov09}
\bibinfo{author}{\bibfnamefont{A.~R.} \bibnamefont{Akhmerov}}
  \bibnamefont{et~al.}, \bibinfo{journal}{Phys. Rev. B}
  \textbf{\bibinfo{volume}{80}}, \bibinfo{pages}{195320}
  (\bibinfo{year}{2009}).

\bibitem[{\citenamefont{Fisher and Lee}(1981)}]{FisherLee81}
\bibinfo{author}{\bibfnamefont{D.~S.} \bibnamefont{Fisher}} \bibnamefont{and}
  \bibinfo{author}{\bibfnamefont{P.~A.} \bibnamefont{Lee}},
  \bibinfo{journal}{Phys. Rev. B} \textbf{\bibinfo{volume}{23}},
  \bibinfo{pages}{6851} (\bibinfo{year}{1981}).

\end{thebibliography}

\end{document}